\documentclass[a4paper,11pt]{article}
\usepackage{jinstpub} 
\usepackage{lineno}
\usepackage{comment}

\title{\boldmath Optimizing the tuning range of the Fermilab Recycler 53MHz RF cavities by exploiting the avoided crossing of the tuner and cavity modes.}

\author{A. E. Roger $^a$, R. Ainsworth $^b$}
\affiliation{$^a$ University of Cambridge,\\
Cambridge, UK}
\affiliation{$^b$ Fermi National Laboratory,\\
Batavia, Illinois, US}

\emailAdd{ar2118@cam.ac.uk}

\abstract{The electromagnetic eigenmodes of the Fermilab Recycler RF cavities were simulated with ACE3P running on NERSC's `Perlmutter' supercomputer in an attempt to optimize their tuning range to comfortably support slipstacking. A theory, based on coupled circuits and avoided crossings, was developed to explain the origins of the tuning range for tuner-cavity waveguide systems. This provided a logical, rigorous method to carry out this optimization, although engineering problems due to high voltages in the tuner and the operational $\mu$ range of the garnet had to be carefully implemented. It was shown that the desired goal of 10kHz of tuning was not attainable with the current cavity specifications. The maximum tuning range obtained was 5.59kHz for a single tuner with a coaxial line of length 63in. A double tuner cavity was also studied using the simulation software, although with little improvement in the tuning range. A promising technique to improve the tuning range involving the desynchronization of the garnets' magnetic permeabilities is proposed.}


\begin{document}
\maketitle
\flushbottom

\section{Introduction}

The Fermi National Laboratory (FNAL) Accelerator Complex has recently undergone upgrades to increase proton output from the Main Injector (MI) to 120 GeV for the protons on target for the new neutrino experiment NO$\nu$A, which studies neutrino oscillations ($\nu_\mu \longrightarrow \nu_e$). Until May 2012, injection, slip-stacking and acceleration was all performed in the MI. Now, injection and slip-stacking will occur in the Recycler Ring (RR) in order to increase beam power.

Slip-stacking is a process developed by CERN by which two beams are merged or `stacked' within each other. This is achieved by slowing down one beam and injecting the other, all the while keeping them in phase to finally revert both to the same energy.

As such, the 53 MHz RF cavities are needed for slip-stacking in the Fermilab Recycler \cite{slipstacking}. The current tuning range (the maximmum frequency of the accelerating mode) is ~3kHz. This is enough for slip-stacking which only needs 1260Hz of tuning, but an actual range of $\sim$10kHz would be ideal to account for frequency drift due to heating.

The recycler RF cavities consist of a two large coaxial cylinders shorted at one end and with a capacitive gap at the other. The gap allows for a strong electric field that can accelerate charged particles. The cavity operates in the TEM mode (remaining comfortably below the TE11 mode). The tuner, located near the shorted end of the cavity, is coupled with the cavity using a coupling loop. The tuner is a solid copper coaxial line (for this cavity, a 50$\Omega$ 3$\frac{1}{8}$in coaxial line) with a ferrite garnet connected at the end. It also operates in the TEM mode well below TE11. A D.C. current is applied to a solenoid surrounding the garnet to change its magnetic permeability which in turn changes the frequency of the accelerating mode of the cavity. Unlike most parallel-biased cavities, the bias field for this cavity is perpendicular to the RF magnetic field. 

It is difficult to calculate the frequency shift partially due to the uncertainty on the magnetic field in the garnet, and also due to lack of published data on the properties of biased garnet. Here, we will attempt to describe this behaviour and understand more about the limitations of the tuning range. To do this, the cavity-tuner system was modelled and electromagnetic simulations were carried out to investigate the dependence of the accelerating frequency on different variables. In addition, simulations were done on a double tuner to propose another way to increase the tuning range in the recycler.

\label{sec:intro}

\section{Software}
Advanced Computational Electromagnetics 3D Parallel (ACE3P), developed by the SLAC National Laboratory, is a set of parallel finite-element multi-physics simulation softwares for electromagnetic, thermal and mechanical modelling purposes. Omega3P was used for this investigation: a module of ACE3P used to simulate electromagnetic eigenmodes in lossy or externally loaded systems (i.e. with boundary conditions) in the frequency domain.

\begin{figure}[htbp]
\centering
\includegraphics[width=.45\textwidth]{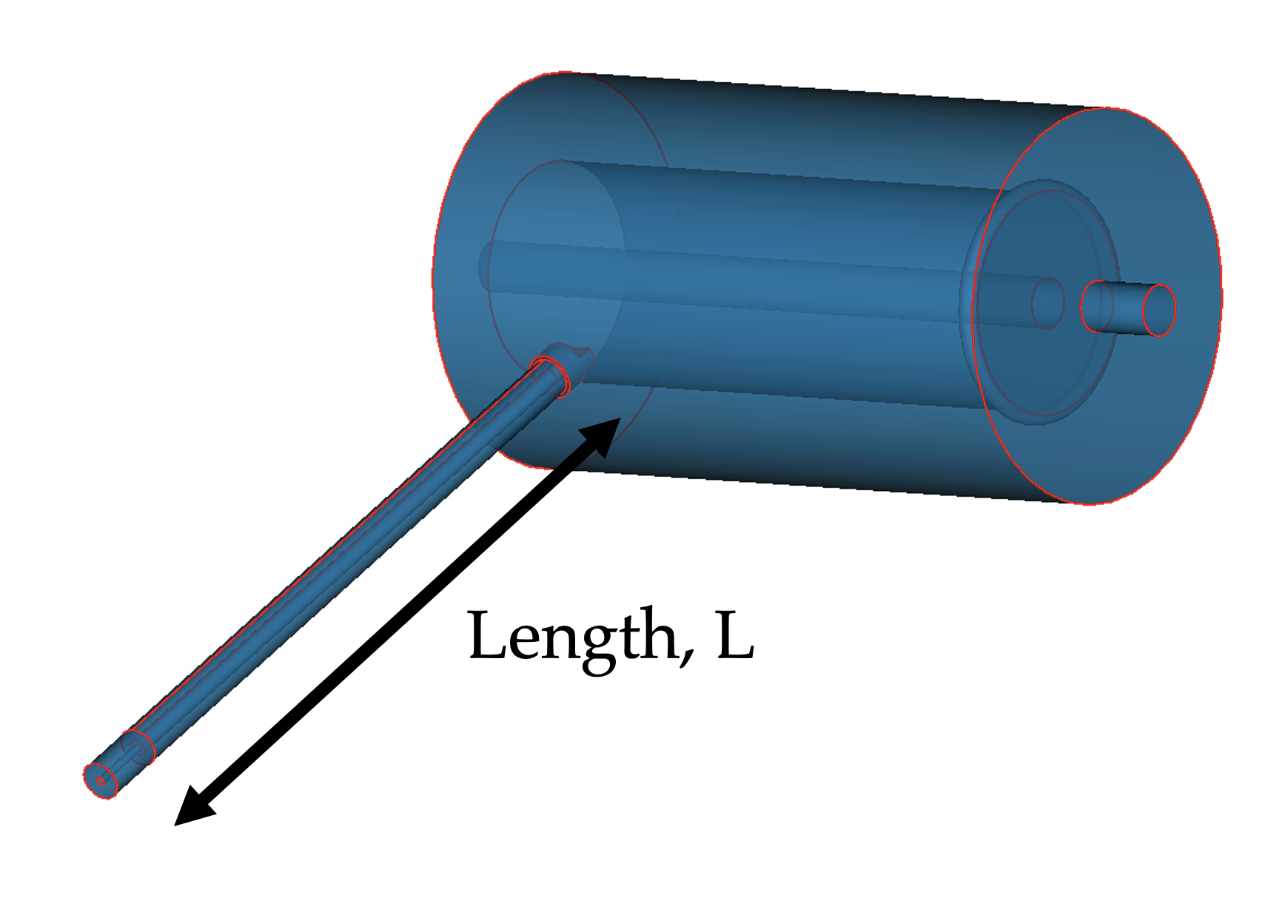}
\qquad
\hspace{-1cm}
\includegraphics[width=.5\textwidth]{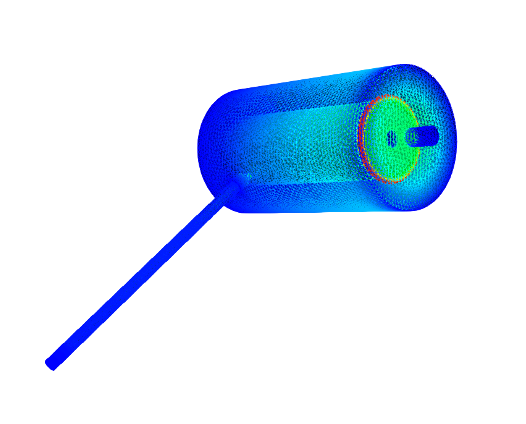}
\caption{Model of the Fermilab RR RF cavity in Cubit (left), and its rendering in Paraview after eigenmode analysis (right). The colours indicate the strength of the electric field. As expected, the field is strongest near the gap and weakest near the shorted end.\label{fig:a}}
\end{figure}

\begin{figure}[htbp]
    \centering
    \includegraphics[width=.3\textwidth]{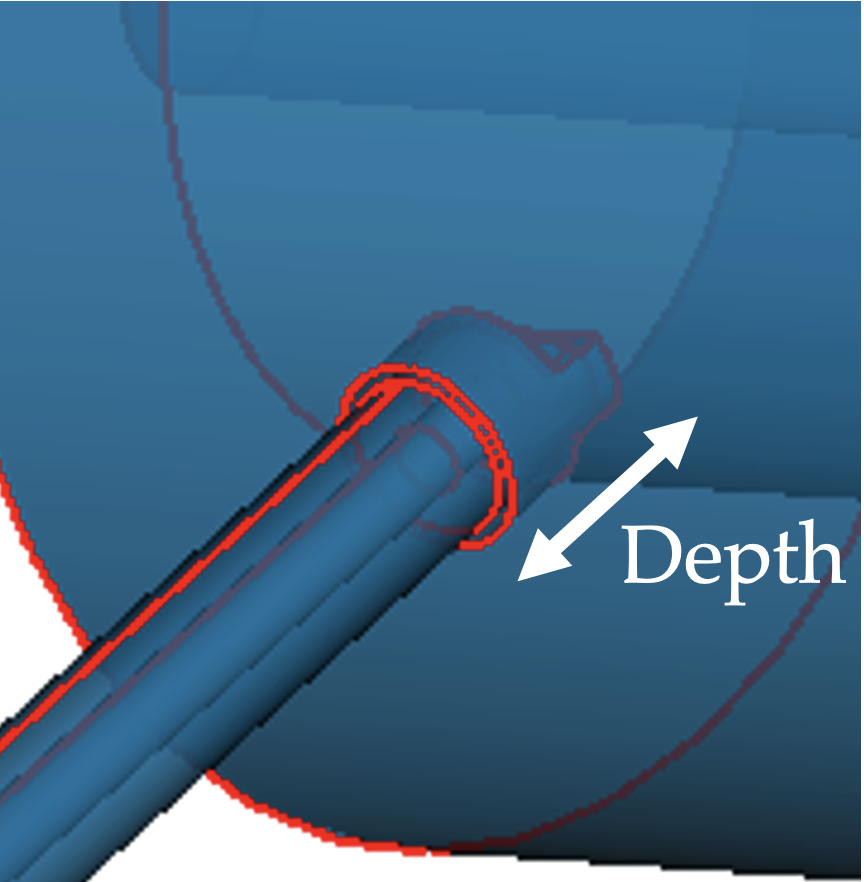}
    \qquad
    \includegraphics[width=.6\textwidth]{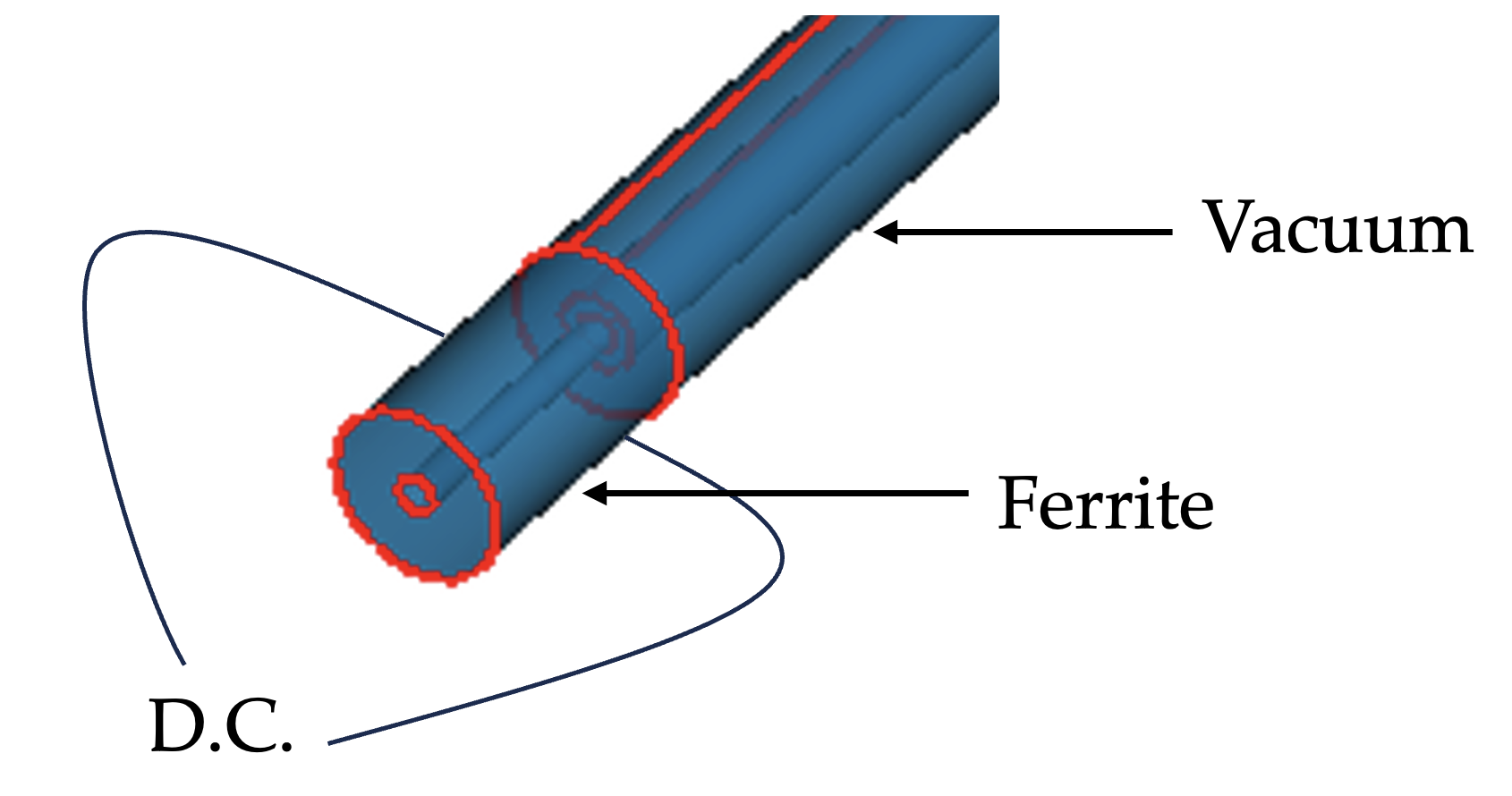}
    \caption{Images of the coupling loop (left) and ferrite garnet (right) in Cubit. The D.C. current changes the magnetic permeability of the ferrite, $\mu_\text{ferr}$. Methods exist to obtain the approximate value of $\mu_\text{ferr}$ as a function of current \cite{robyn}. The inner radius of the ferrite garnet is smaller than that of the coaxial line. This is done to keep the impedance of the garnet at $\sim$50$\Omega$ during operation to match the rest of the line.\label{fig:b}}
    \end{figure}

The model was first created in Cubit (see Figure \ref{fig:a} and \ref{fig:b}) \cite{cubit} and Omega3P was run on the NERSC `Perlmutter' Supercomputer in Berkeley. The system was analysed and the eigenmodes were detected. Consequently, information about these modes was gathered (e.g. frequency of the mode, quality factor, the total energy of the mode (normalised)...). The individual modes could then be post-processed to gather more information such as the R/Q factor or the electric field at specific points for that mode.

Each eigenmode could be rendered on the model in a second software, Paraview \cite{paraview} (see Figure \ref{fig:a}). These renderings provided a useful visual representation of the modes with which we could observe the electric field, magnetic field and more.

\section{Avoided crossing theory}
\begin{figure}[htbp]
    \centering
    \includegraphics[width=0.76\textwidth]{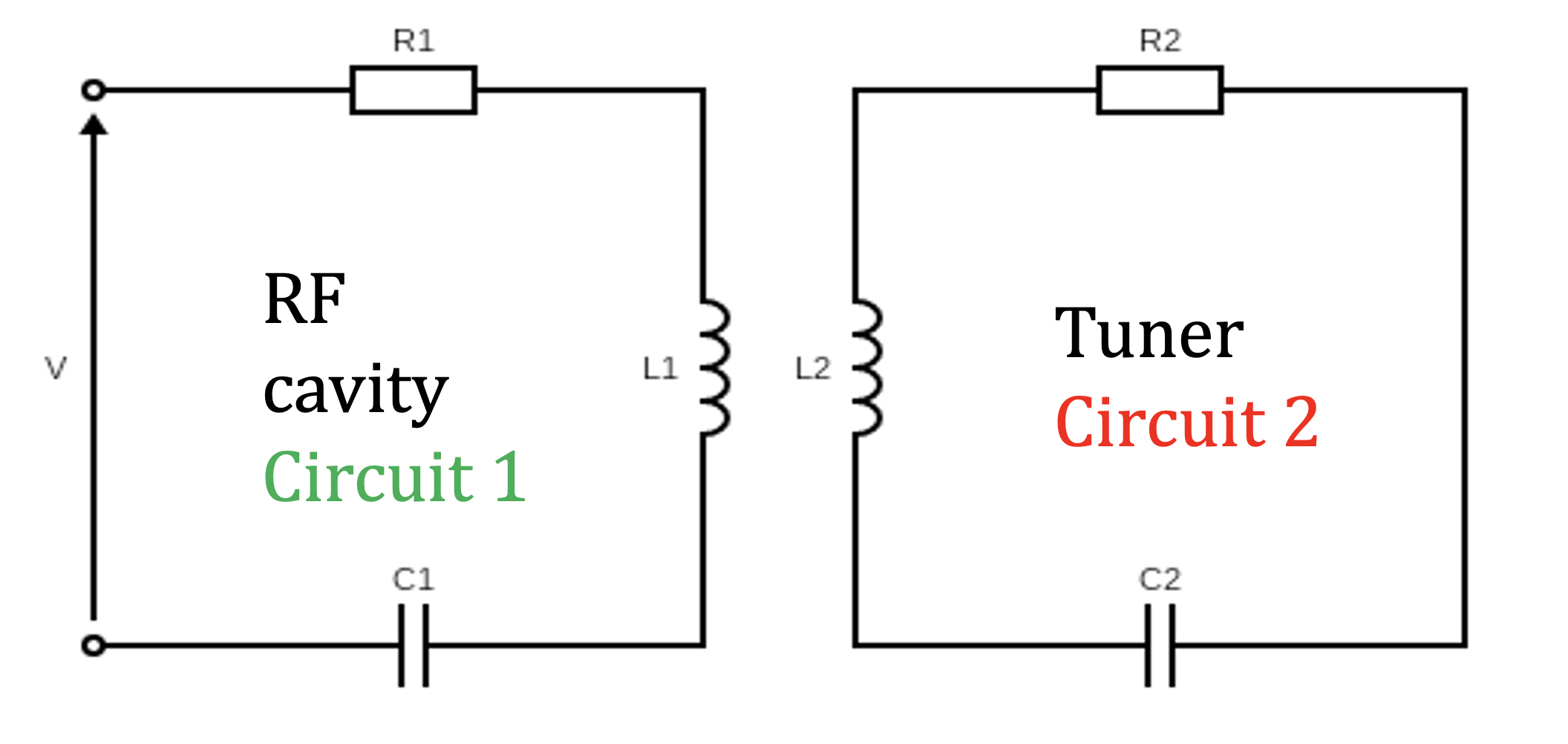}
    \caption{Effective circuit of our tuner and RF cavity. The two circuits are coupled together via their inductors. The coupling constant $k$ will exist in the range $0<k<1$. The voltage $V$ is alternating.\label{fig:c}}
    \end{figure}

The behaviour of the accelerating mode of the cavity can be predicted using an equivalent transmission line circuit for the cavity-tuner system (see Figure \ref{fig:c}). The circuit chosen was two coupled LCR circuits, where the coupling of the inductors corresponds to the coupling of the tuner and cavity's magnetic fields via the coupling loop.  

Let circuit 1 represent the cavity and circuit 2 represent the tuner, with $M$ being the mutual inductance. We can solve the circuit using impedance considerations, calling the impedances of the cavity and tuner $Z_1$ and $Z_2$ respectively:
\begin{align}
	Z_1 (\omega) = R_1 + jL_1 \omega - \frac{j}{C_1 \omega}
	\\
	Z_2 (\omega )= R_2 + jL_2 \omega - \frac{j}{C_2 \omega}
\end{align}

Let us suppose solutions of the form,
\begin{align*}
	I_1 (t) &= i_1 e^{j \omega t}
	\\
	I_2 (t) &= i_2 e^{j \omega t}, \text{ where } I_1, I_2 \in \mathbb{C}
\end{align*}

From Kirchoff's laws,
\begin{align}
	V = Z_1 i_1 + j M \omega i_2 \text{ (circuit 1)}
	\\
	0 = Z_2 i_2 + j M \omega i_1 \text{ (circuit 2)}
\end{align}

Rearranging:
\begin{equation}
	{i_1(\omega) = \frac{V Z_2(\omega)}{Z_1(\omega) Z_2(\omega) + M^2 \omega^2}}
\end{equation}

\begin{figure}[h]
    \centering
    \includegraphics[width=.47\textwidth]{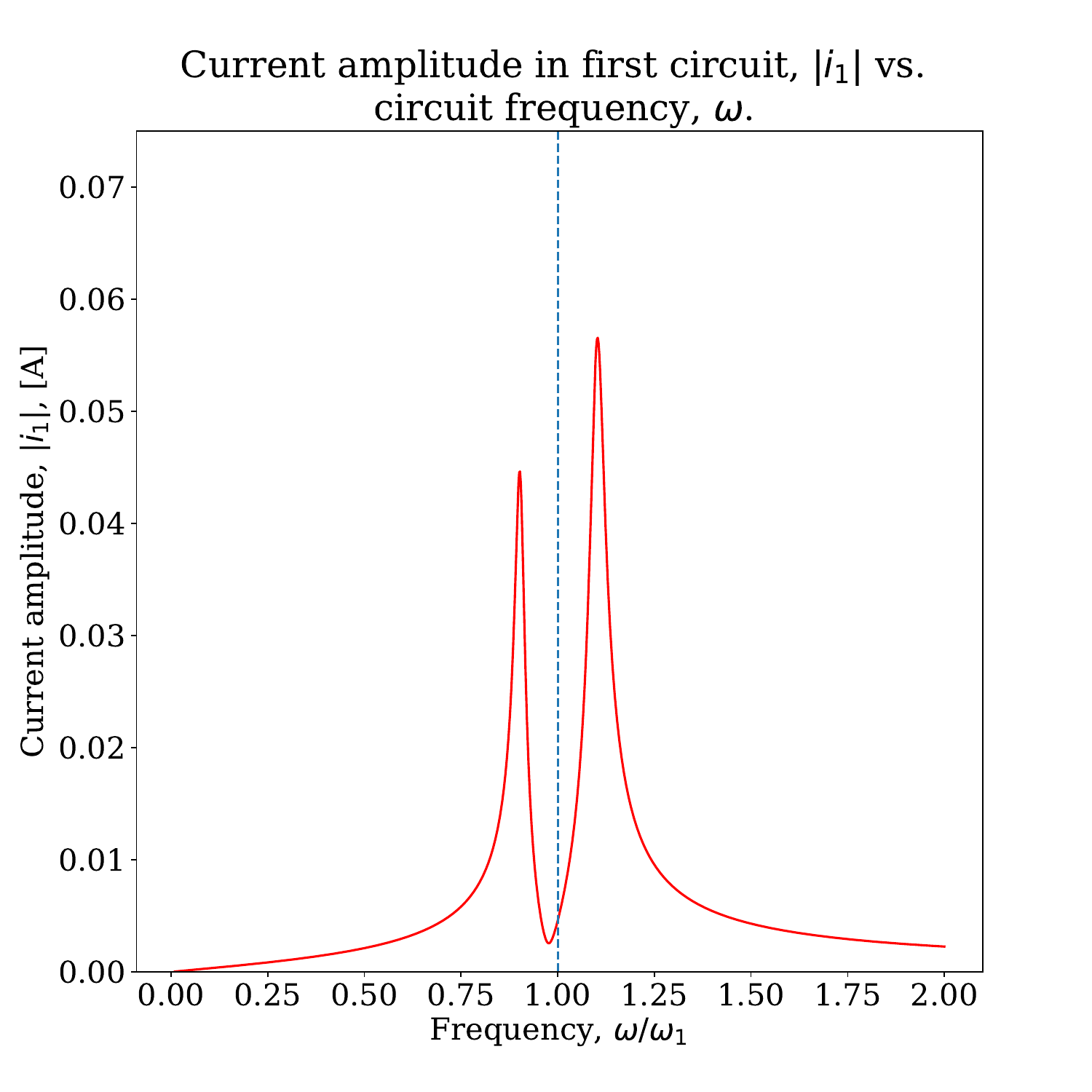}
    \qquad
    \includegraphics[width=.47\textwidth]{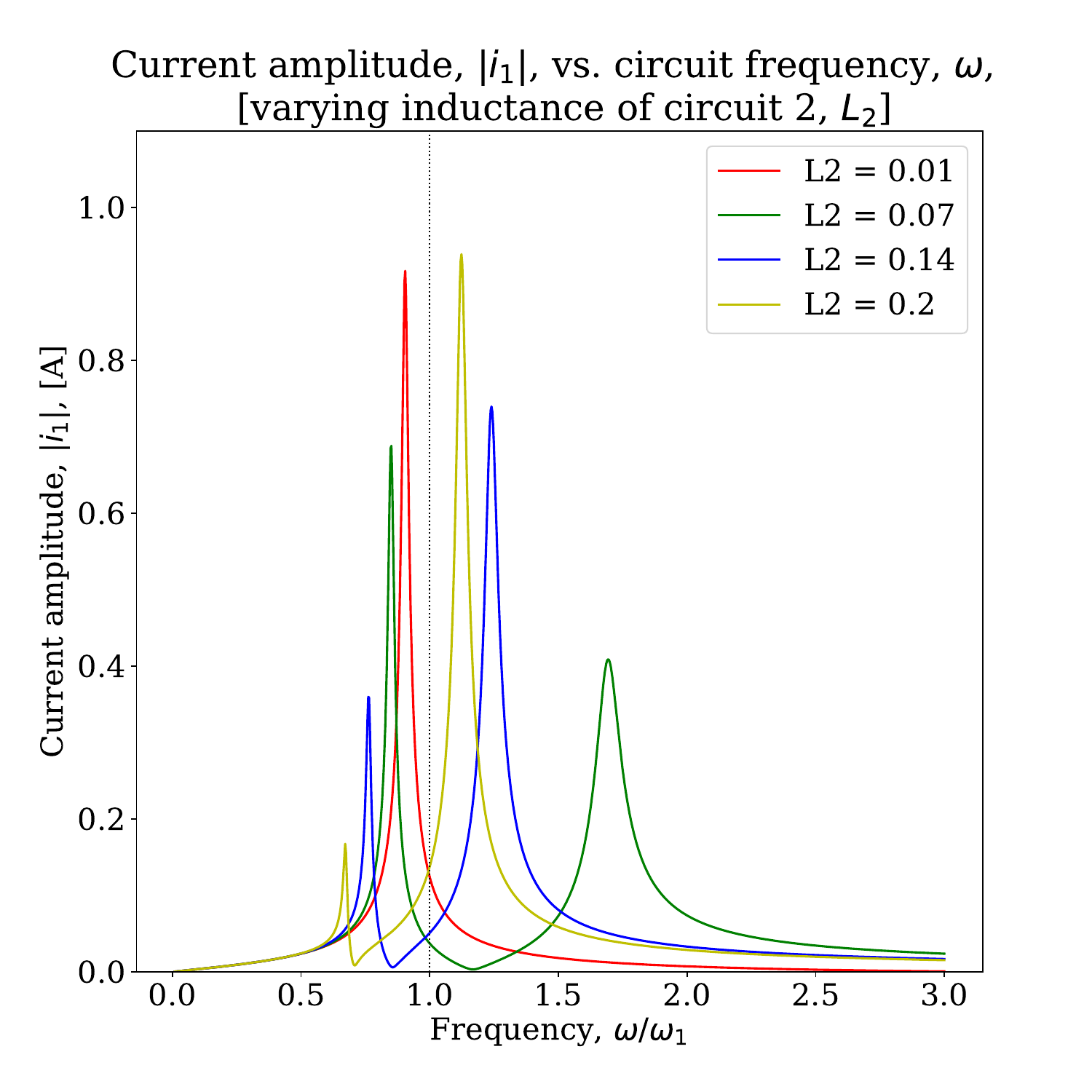}
    \caption{Graphs displaying the coupling behaviour between the tuner and RF cavity. The response function always displays two peaks on either side of circuit 1's characteristic frequency ($\omega_1$), corresponding to two `circuit modes' and hence the first two modes of the cavity. For these plots values for the components of the circuit were chosen to suitably display the behaviour of the system ($L_1 = 0.1 \text{H}, C_1 = 1 \mu \text{F}, R_1 = 10 \Omega, C_2 = 1 \mu \text{F}, R_2 = 10 \Omega, M = 0.02$).\label{fig:d}}
    \end{figure}

We can plot the response function of $|i_1|$ as a function of circuit frequency $\omega$ (see Figure \ref{fig:d}): two peaks are discernible. As a large current corresponds to a large power dissipated into the circuit, we use the analogy that a peak in current $|i_1|$ corresponds to a mode in the cavity. This follows because energy in the cavity `wants' to be dissipated most when it is operating in its electromagnetic modes.

Consequently, the frequency at which a peak in $|i_1|$ exists effectively corresponds to the frequency of a mode in the cavity - desired for the acceleration of particles. 

One issue that arises from this circuit analogy is the lack of higher order modes predicted by the circuit:

In most waveguides, the electromagnetic field is almost completely characterized by the first dominant mode. In fact this is the case for our system as the tuner and cavity are operating in TEM modes well below the next highest mode (TE11). Hence, we can safely neglect the effect of higher order modes. After making this assumption, we can notice that the modulation of the electric/magnetic field with position is closely suggestive of voltage/current behavior in a transmission line. It is thereby implied that the electromagnetic fields almost everywhere can be completely represented as an equivalent transmission line circuit specific to the RF system and its geometry. This outlines a strong argument for using the equivalent transmission line circuit above.

If one wished to simulate these non-propagating modes, more complex transmission lines circuits would need to be used, with each new circuit corresponding to one mode. Interestingly, the complete behaviour of an electromagnetic field can be characterized by an infinite number of transmission line circuits \cite{handbook}. Small kinks in the solid and discontinuities may cause these higher order modes to propagate - for example some might exists near the coupling loop - but in this investigation focus remained on dominant modes, deeming this issue to be negligible. 

Now consider circuit 2. For a coaxial line (of inner and outer radii $a$ and $b$ respectively) the inductance per unit length is $L = \frac{\mu \mu_0}{2\pi} \ln{\left(\frac{a}{b}\right)}$, hence we notice that the inductance $L_2$ is directly proportional to both the overall magnetic permeability $\mu$ of the tuner and its length $l$. The magnetic permeability and inner radius are not uniform along the length of the tuner (because of the ferrite garnet), but the analysis remains true if we consider the effective $\mu_{\text{eff}}$ and length $l_\text{eff}$ of the tuner. 

\begin{figure}[h]
    \centering
    \includegraphics[width=.47\textwidth]{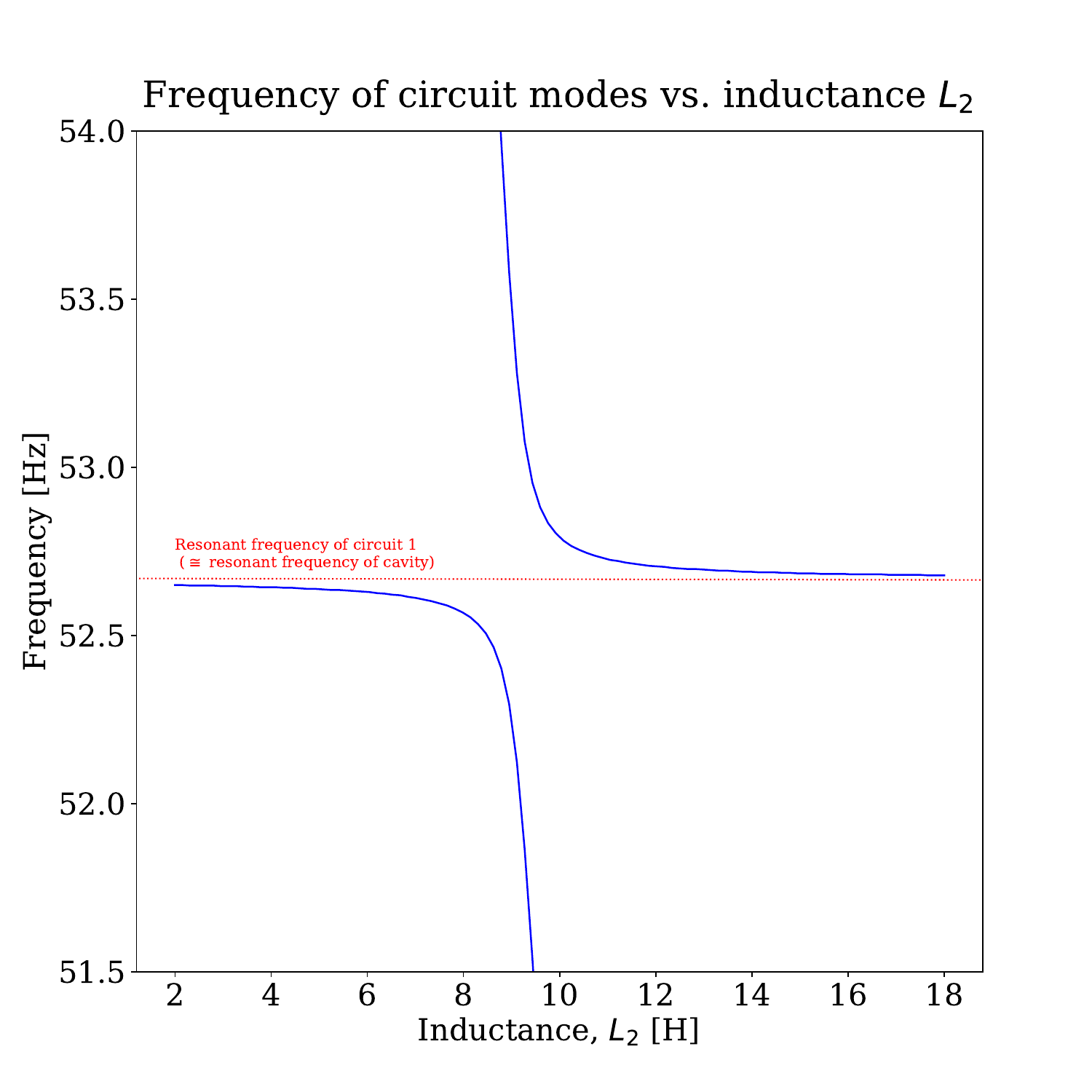}
    \qquad
    \includegraphics[width=.47\textwidth]{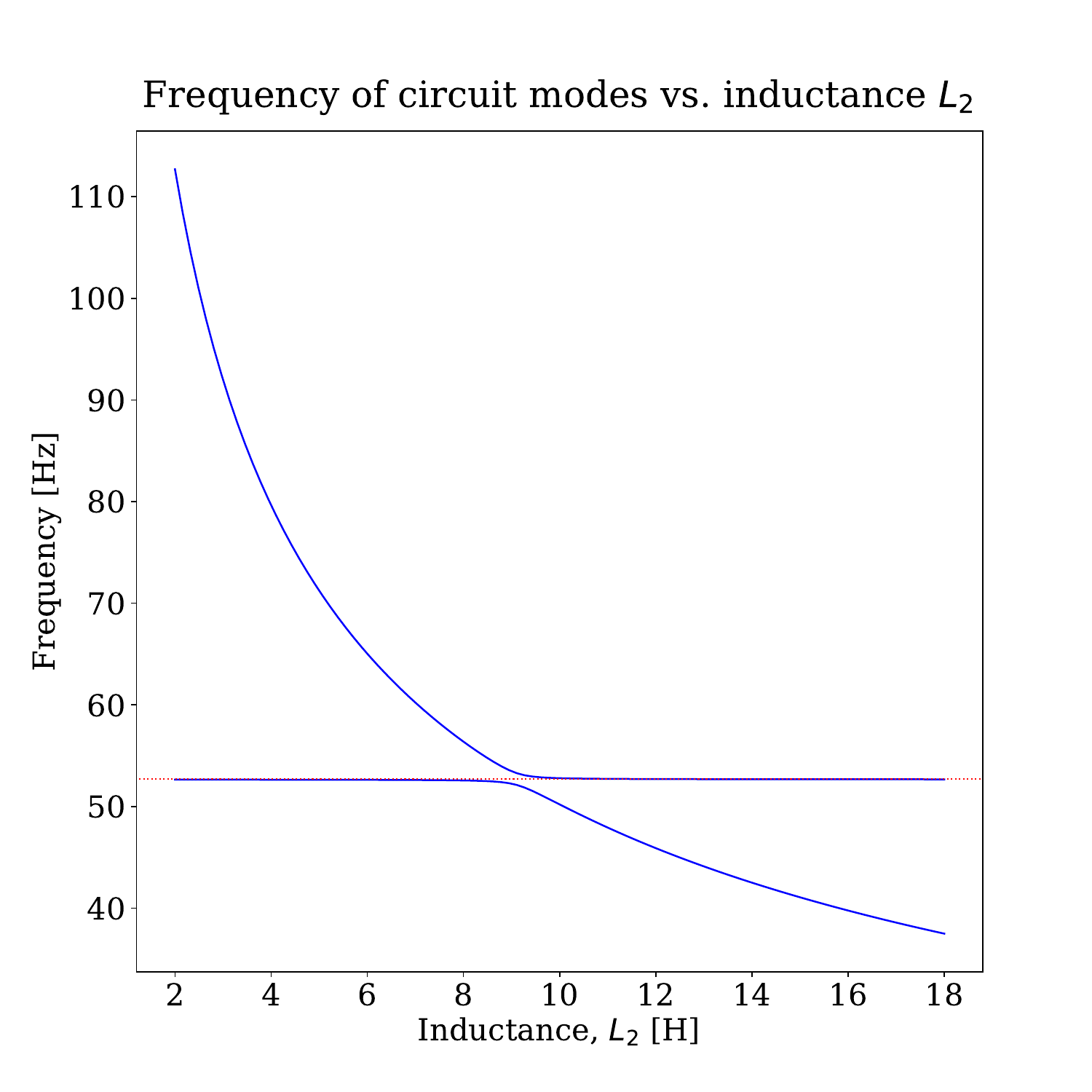}
    \qquad
\caption{Graphs simulating the coupling behaviour between the tuner and RF cavity using the effective circuit analogy. Away from the crossing, the flat regions correspond almost completely to a cavity mode whilst the curved regions corresponds to that of the tuner. Closer to the crossing the electromagnetic modes become shared between the tuner and the cavity. Values for the components of the circuit were chosen to conveniently mimic the resonant frequency of the actual cavity in question ($L_1 = 9.129 \text{H}, C_1 = 1 \mu \text{F}, R_1 = 0.2 \Omega, C_2 = 1 \mu F, R_2 = 0.2 \Omega, M = 0.2$). \label{fig:tunermodes_theory}}
\end{figure}

We can then claim that changing $L_2$ corresponds to changing $\mu_{\text{eff}}$ or $l_\text{eff}$. The location of the two $|i_1|$ peaks in $\omega$ space changes with $L_2$. If we plot the frequency of the two peaks for many values of $L_2$, we notice that there are actually two `circuit' modes that avoid each other at some $L_2$. After running simulations and examining the field in Paraview, it became clear that the mode which curves down the frequency plot (see Figure \ref{fig:tunermodes_theory}) actually corresponds to a mode existing largely in the tuner. As expected, the mode centred around the accelerating frequency of the cavity corresponds to a mode existing in the cavity. 

Using this analogy, we predict that the frequency of the cavity mode changes solely due to mode coupling with the tuner which usefully creates an avoided crossing. This phenomenon is seen in many other fields of physics ranging from mechanics to superconducting quantum interference devices \cite{nature}. With this idea, it seems that we should aim to vary the inductance of the tuner (and hence $\mu_\text{ferr}$) to get as close as possible to the avoided crossing where the frequency decreases faster and faster. However, the amplitude of the tuner mode significantly increases in this region which causes various other problems that will be explained in section \ref{sec:limitations}.

\begin{figure}[h]
    \includegraphics[width=.47\textwidth]{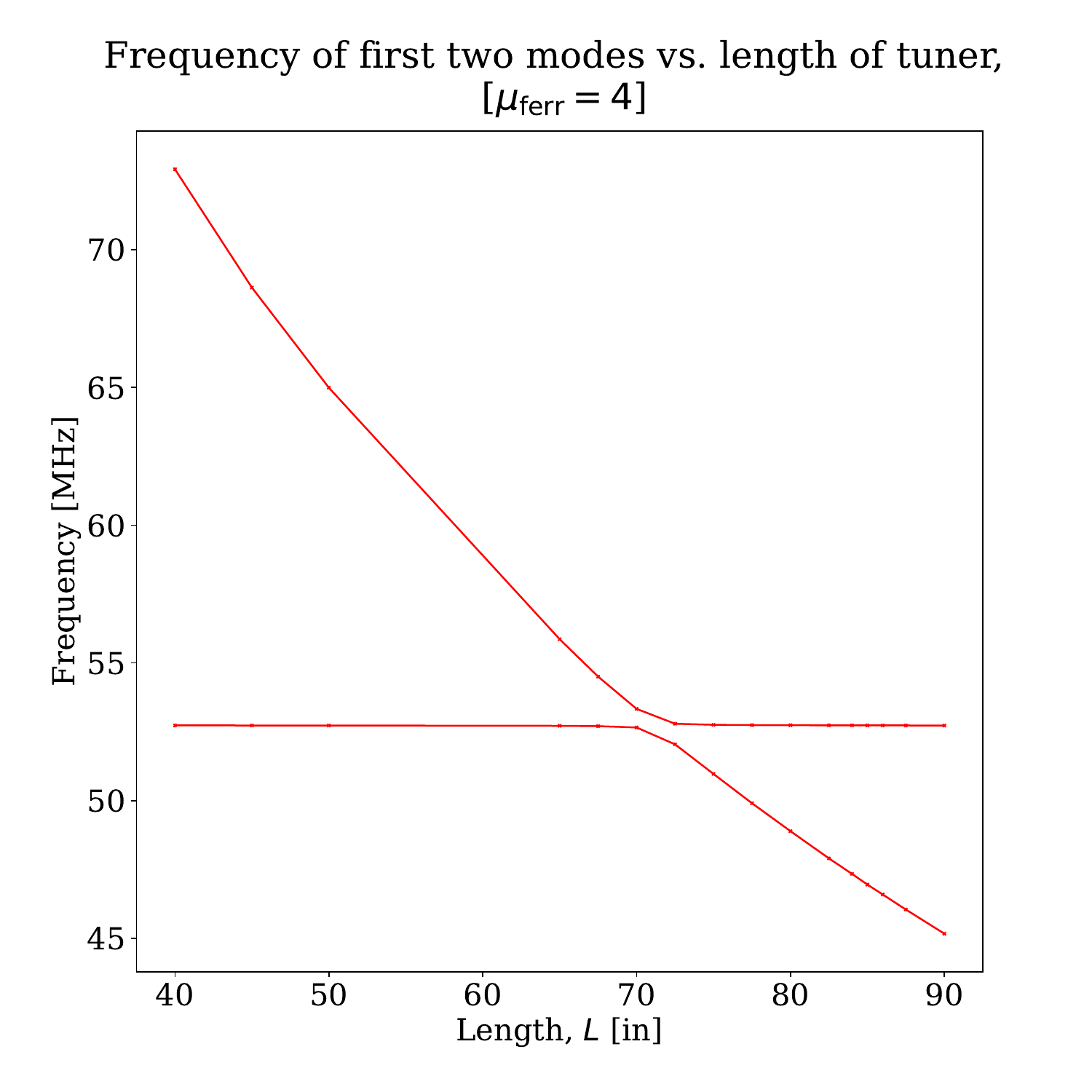}
    \qquad
    \includegraphics[width=.47\textwidth]{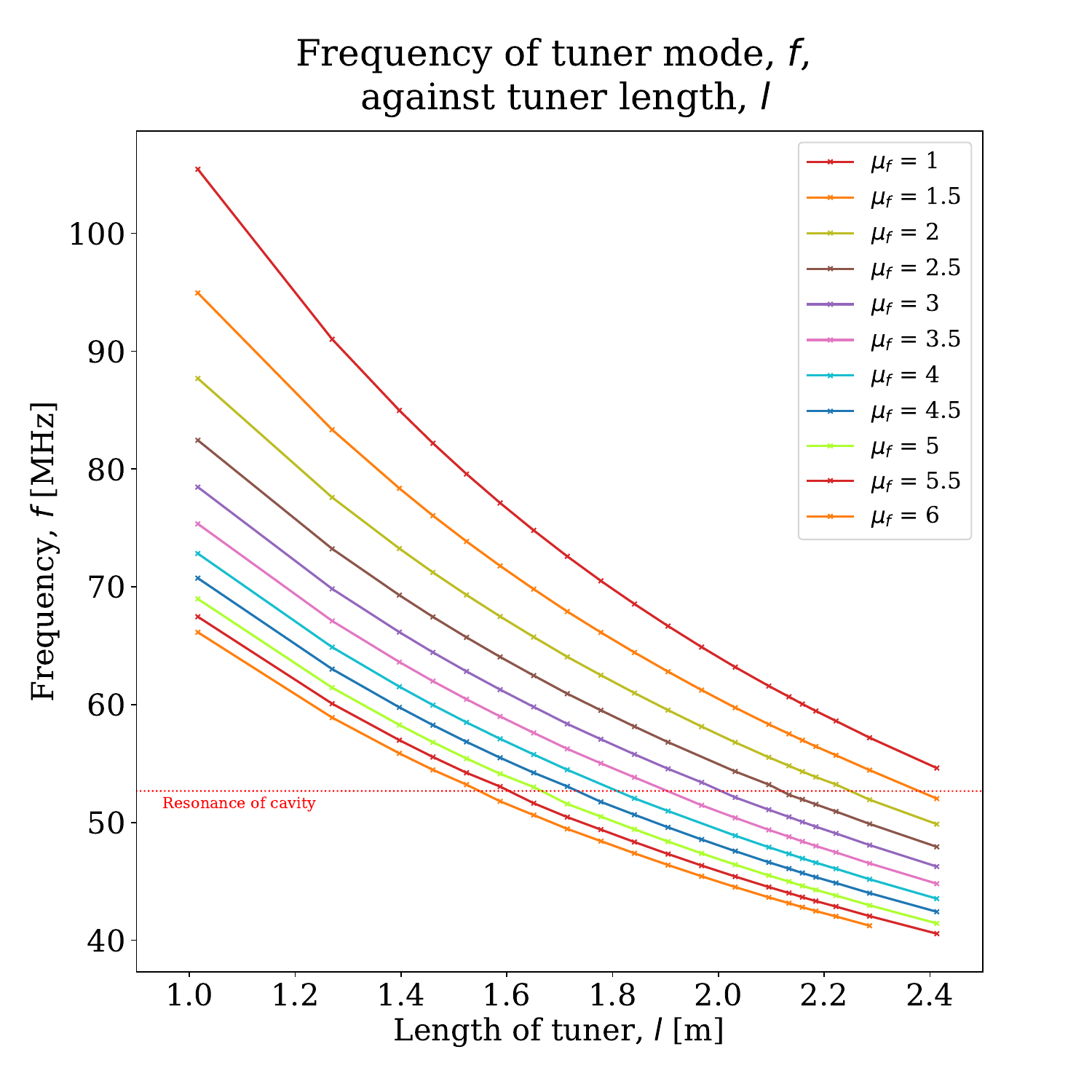}
\caption{Graphs of actual data gathered from simulations on Ace3P. We can compare the eigenmode graph (left) with Figure \ref{fig:tunermodes_theory} because (as shown earlier) the inductance of the tuner will increase with increasing tuner length ($\propto l_\text{eff}$). The avoided crossing is evident. For the plot on the right, the avoided crossing was ignored and the curves were extrapolated across it to display the overall behaviour of the tuner modes (attempting to represent the behaviour as if the cavity were not there). This mode is further analysed in section \ref{sec:tunermode}. \label{fig:tunermodes_data}}
\end{figure}

\section{Simulations}

\subsection{Data vs. Theory}

The frequency of the first two modes of the cavity-tuner system were recorded as both the magnetic permeability of the ferrite, $\mu_\text{ferr}$ and the length of the tuner coaxial line, $l$ were varied. Mode coupling behavior was observed from the data in both cases, and an avoided crossing was always discernible for specific values of $l$ or $\mu_\text{ferr}$. After choosing convenient values for the equivalent circuit components, the similarity between numerical and theoretical plots was striking (see Figure \ref{fig:tunermodes_data}).

\subsection{Analysis of the tuner mode}
\label{sec:tunermode}

The mode in the tuner was of key interest as it causes the avoided crossing. Understanding more about this mode would allow us to predict where this crossing occurs in $(\mu_\text{ferr}, l)$ space and hence choose our length of tuner and operational $\mu_\text{ferr}$ range to get as much frequency shift as possible. As previously mentioned, the tuner operates in the TEM mode well below the TE11 frequency ($f_\text{TE11}=$236MHz for this tuner). For this mode, the tuner acts as a cavity resonator \cite{pozar}. 

The ferrite garnet is shorted so the $\vec{E}$ standing wave must be zero at this end. Finding an analytical expression for this wave would be complex as it is difficult to know where the second reflection is (it seems to be within the cavity, let us call the distance between this `bounce' and the tuner-cavity interface $\delta$): solving Maxwell's equations for the full system is likely to be the only way to obtain an analytical expression for this mode. An approximate solution can be found by considering the end correction and the time take for the wave to travel back and forth through the tuner. Assuming the fundamental mode within the tuner:
\begin{align}
    T \approx 2 \left( \frac{l_{\text{ferr}} \sqrt{\mu_\text{ferr} \varepsilon_\text{ferr}}}{c} + \frac{l}{c} + \frac{\delta}{c}\right)
\end{align}

Alternatively we can think of $l_{\text{ferr}} \sqrt{\mu_\text{ferr} \varepsilon_\text{ferr}} $ as the effective distance travelled by the wave within the ferrite. We can turn this into the mode frequency:

\begin{align}
    f_\text{parasite} (\mu_\text{ferr}, l) \approx \frac{c}{2(l + l_\text{ferr} \sqrt{\mu_\text{ferr} \varepsilon_\text{ferr}} + \delta )}
\end{align}

\begin{figure}[h]
    \includegraphics[width=1\textwidth]{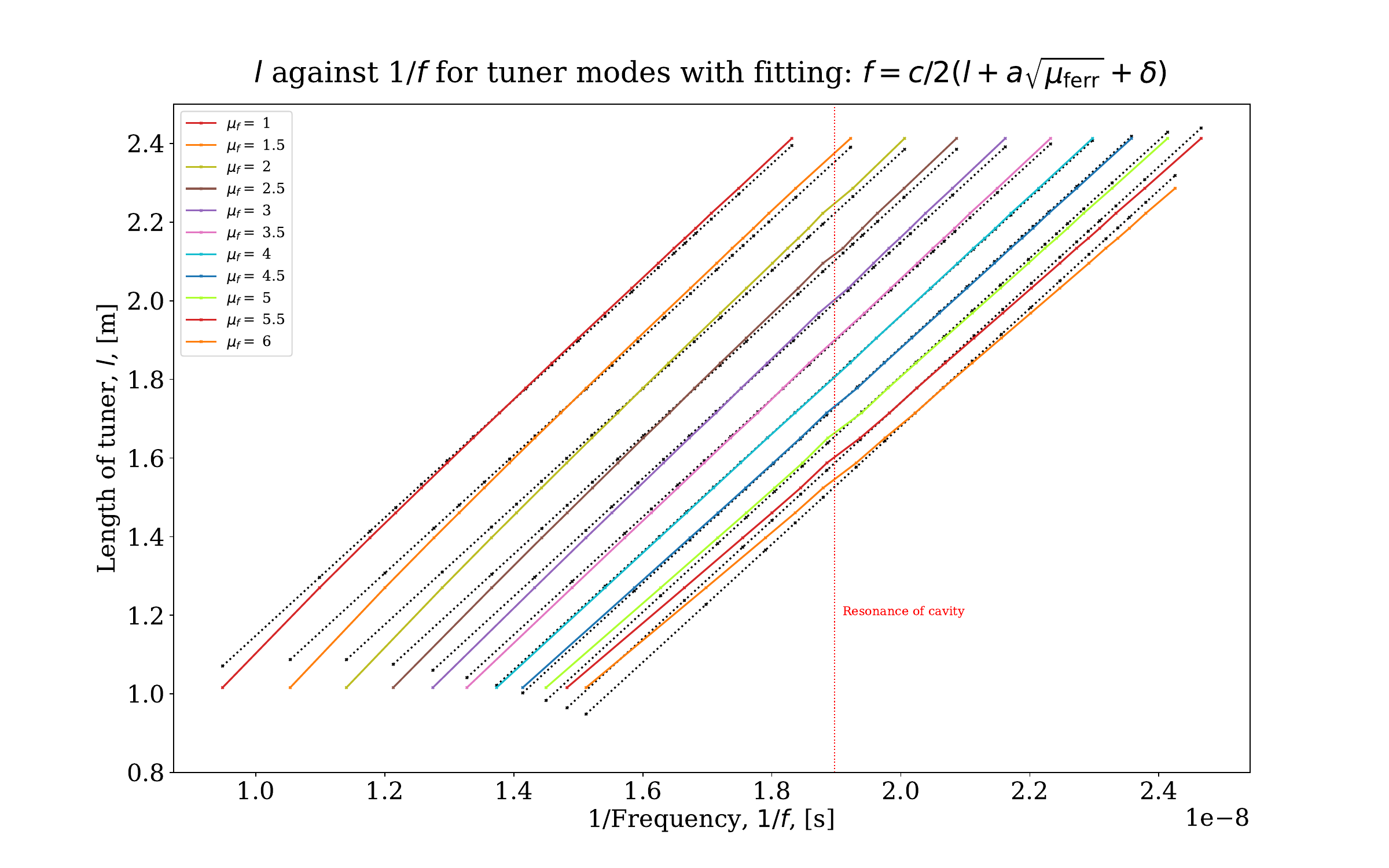}

    \caption{By plotting $l$ against $1/f$ we can linearise the equation found for $f_\text{parasite}$ (4.2). Using a numpy package in python to find the best values of $a$ and $\delta$, the data was well fitted ($a=0.68$ and $\delta=-0.33$).  \label{fig:g}}
    \end{figure}

Adopting the form of this equation to fit the data obtained from simulations, a good match in coefficients was achieved (see Figure \ref{fig:g}). Of course, this equation will only be accurate for a coaxial tuner of these dimensions. Nonetheless, the dependence of the equation on $\mu_\text{ferr}$ and $l$ can be accurately used for other cases if necessary to locate the avoided crossing for any $(\mu_\text{ferr}, l)$.

\section{Limitations of the tuning range}
\label{sec:limitations}

The avoided crossing theory offers an appealing method to optimize the tuning range of an RF cavity. However, approaching this crossing creates engineering issues which needs to be taken into consideration. 

\subsection{Quality factor of the cavity}

As the tuner frequency approaches the cavity resonance frequency, its power flow/amplitude increases significantly. Hence the energy of the mode shifts from being mostly inside the cavity to mostly inside the tuner. Subsequently the quality factor, $Q$, of the mode falls quickly. When designing the cavity, the frequency shift must be such that $Q$ remains above a reasonable value. In practice, this was not a major problem for the RR cavities in question.

\subsection{Voltage breakdown in the tuner}

The cavity-tuner system must operate such that the maximum voltage in the tuner is small enough to prevent breakdown. The peak voltage permitted for this tuner, omitting the safety factor, was calculated to be $13.3$kV \cite{breakdown}.

\begin{figure}[h]
    \includegraphics[width=1.0\textwidth]{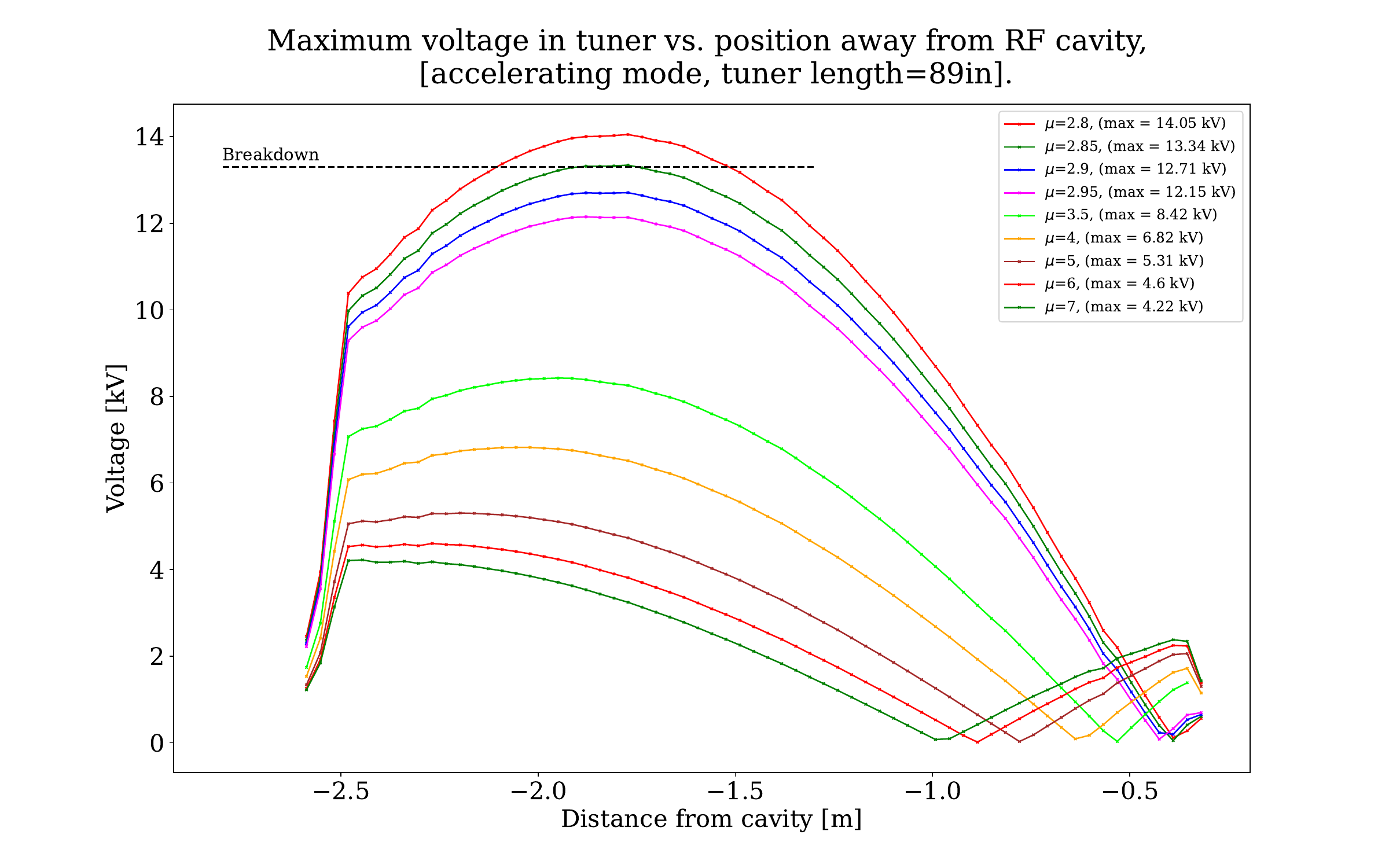}
    \caption{For certain values of $\mu_\text{ferr}$ the voltage surpasses breakdown which is problematic. This corresponds to being too close to the mode crossing, hence restricting the range of frequency attainable by the cavity. A graphical way to optimize the tuning range would be to find the two most extremal voltage curves for some $\mu_\text{ferr}$'s with the upper curve remaining below breakdown (see Section \ref{sec:optimization}). The larger space between voltage lines corresponds to a larger change in frequency. For the tuner length above, the best tuning range attainable (between the two dark green lines, $\mu=7$ to $\mu=2.85$) is 8.40kHz - tightening the $\mu$ range would only decrease this value.\label{fig:breakdown}}
\end{figure}

Maximum voltage, $V_\text{max}$, readings were taken as a function of their position along the tuner (e.g. see Figure \ref{fig:breakdown}). The electric field is strongest at the inner radius of the coaxial line, hence this is where values for the electric field were recorded for our model. These field magnitudes $E_\text{max}$ can be converted to $V_\text{max}$:

Consider the inner cylinder to have charge per length $\lambda$ and the inner and outer radii to be $a$ and $b$ (remembering $V=0$ on the outer radius):

\begin{align}
    E(r) &= \frac{\lambda}{2 \pi \varepsilon_0} \frac{1}{r} \\
    V(r) &= \frac{\lambda}{2 \pi \varepsilon_0} \ln{\left( \frac{b}{r}\right)}\\
    E_\text{max} &= \frac{\lambda}{2 \pi \varepsilon_0} \frac{1}{a}\\
    \implies V_\text{max} &= E_\text{max} \cdot a \ln{\left( \frac{b}{a}\right)}
\end{align}

For the RR cavities, this problem greatly reduced the tuning range attainable by the cavity.

\subsection{Operational mu range of the ferrite garnet}

The operational range of the garnet's magnetic permeability is small for this cavity ($2.5<\mu_\text{ferr}<4$) due to losses at high D.C. current in the ferrite \cite{robyn}. This significantly reduced the tuning range once again.

\section{Final optimization method of the RR cavities}

In summary, the accelerating frequency of the cavity changes the most when it is close to that of the tuner. However, the tuner's breakdown voltage cannot be surpassed and $\mu_\text{ferr}$ must remain between 2.5 and 4. 

By comparing frequency and tuner voltage plots it is clear that the tuner voltage spikes when we approach the avoided crossing. Hence to optimize the tuning range, the length of the tuner must be chosen such that the tuner voltage is maximised (i.e. just under breakdown) for one boundary value of $\mu_\text{ferr}$ (i.e. 2.5 or 4).

\begin{figure}[h]
    \centering
    \includegraphics[width=1\textwidth]{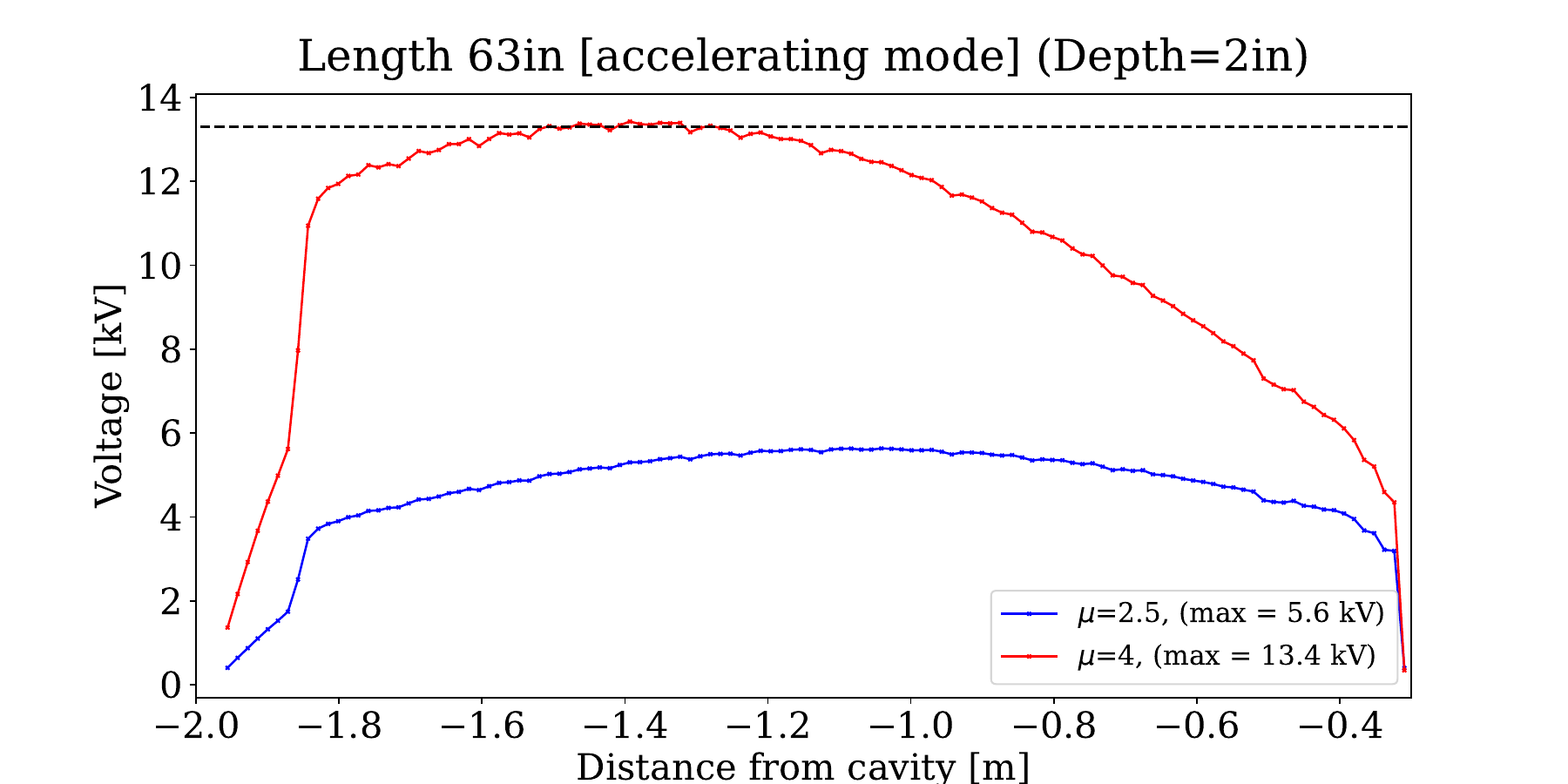}
    \qquad
    \includegraphics[width=1\textwidth]{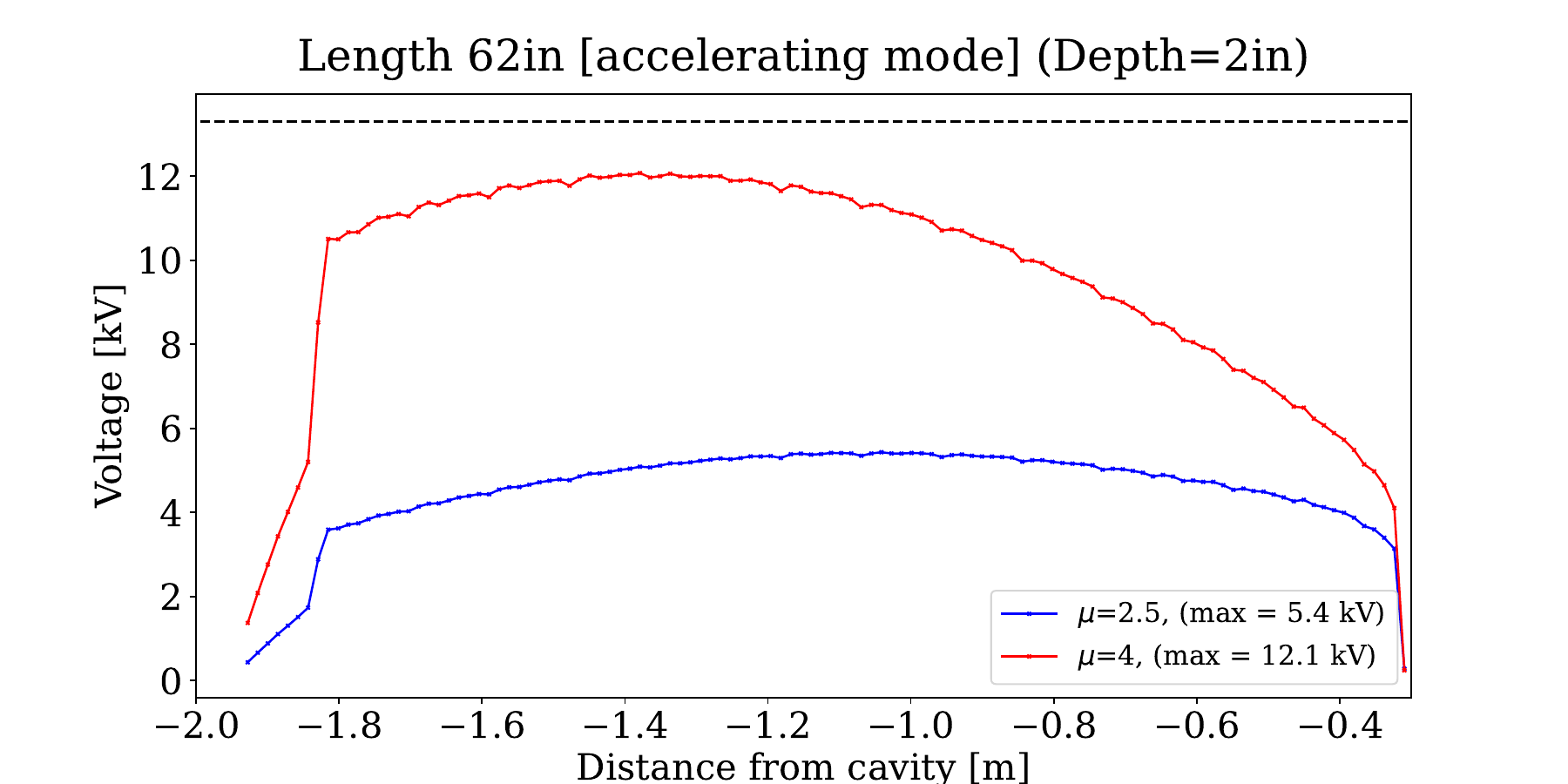}
    \caption{Tuner voltage response for lengths of 63in (optimal, tuning range = 5.59kHz) and 62in (safer, tuning range = 4.85kHz). \label{fig:optimallengths}}
\end{figure}

\begin{figure}[h]
    \centering
    \includegraphics[width=1\textwidth]{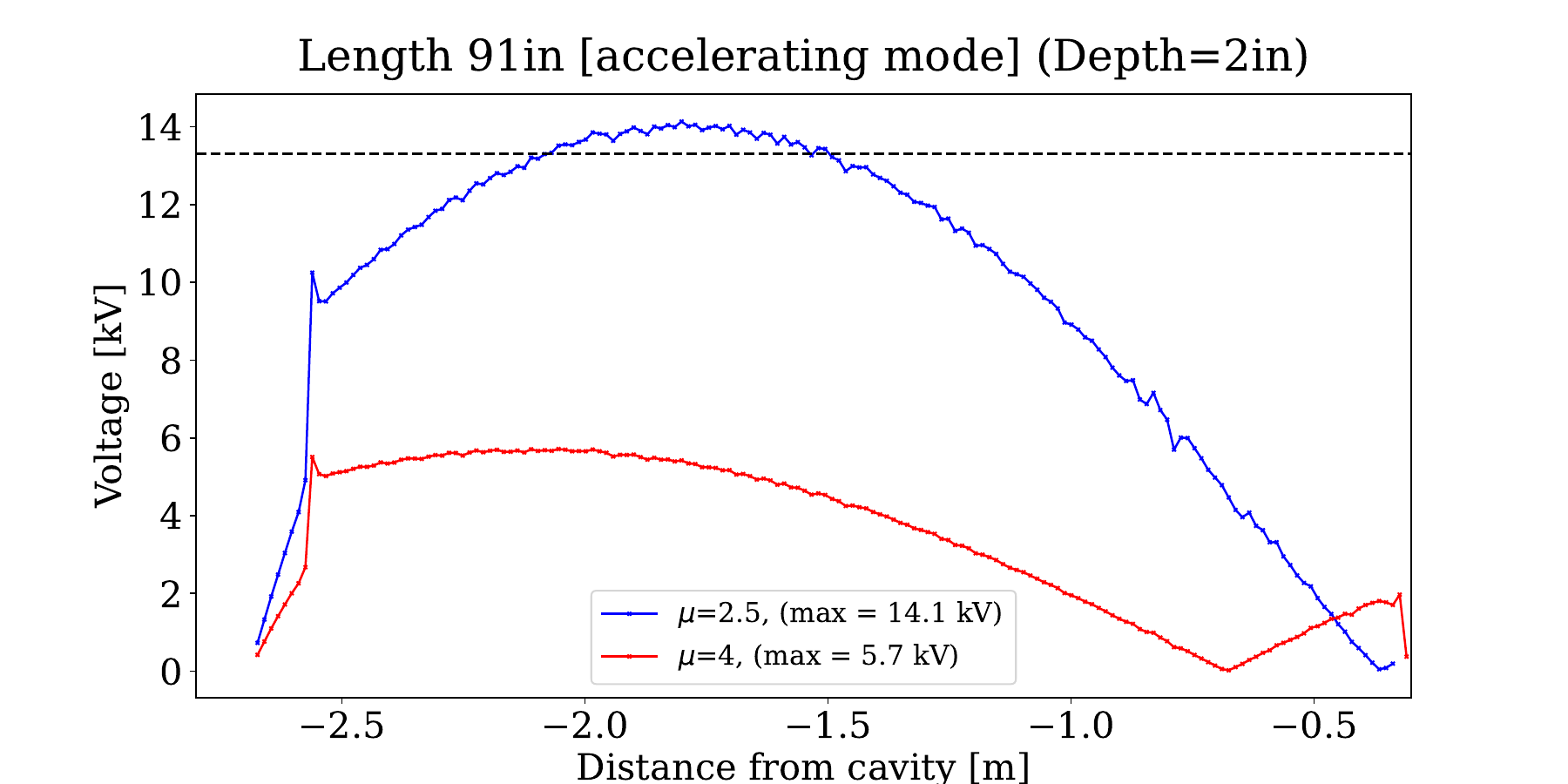}
    \qquad
    \includegraphics[width=1\textwidth]{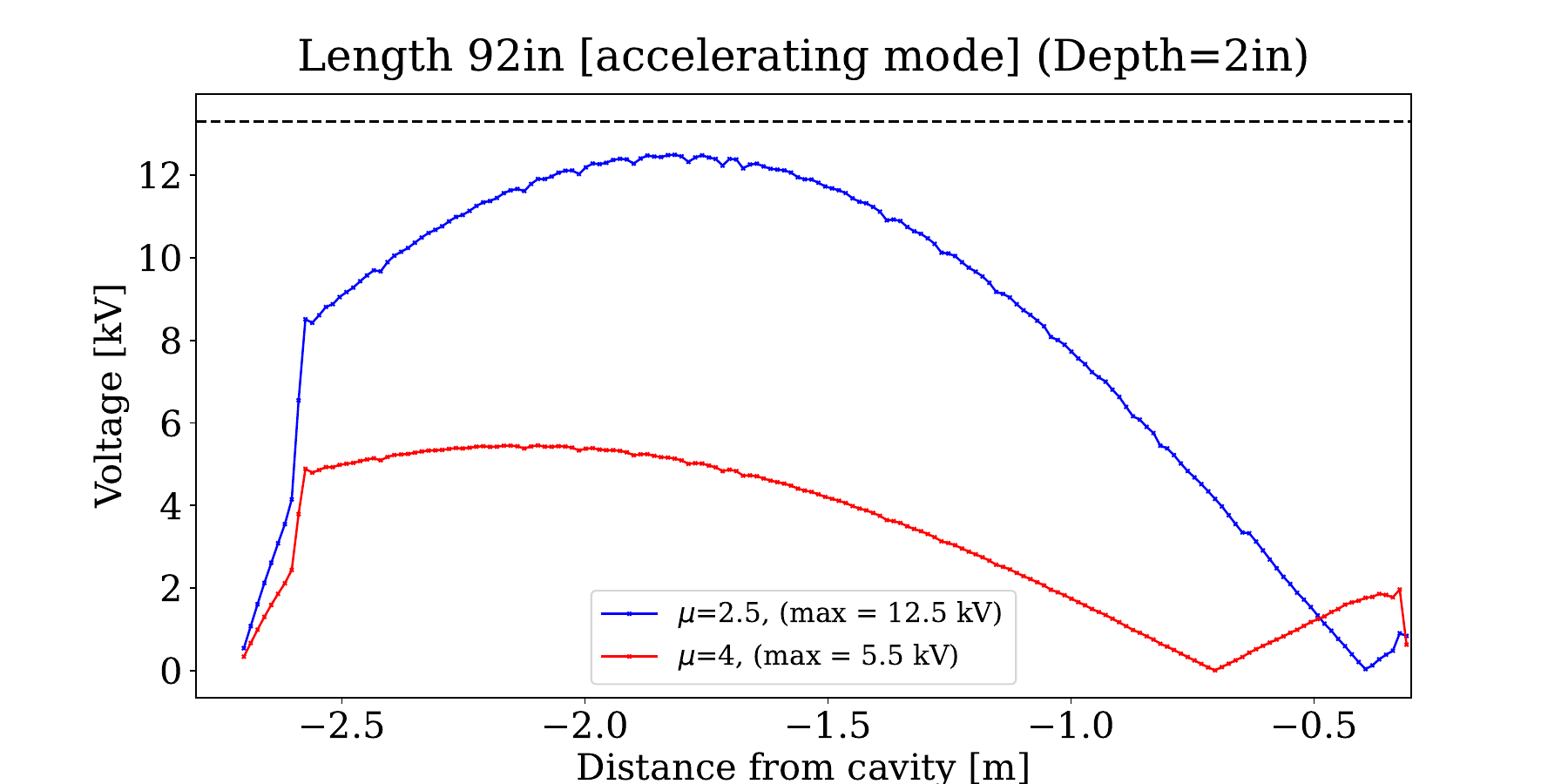}
    \caption{Tuner voltage response for lengths of 91in (second optimal length, tuning range = 5.94kHz) and 92in (safer, tuning range = 5.07kHz). Here, we are on `the other side' of the avoided crossing. Increasing the length is like increasing $\mu_\text{ferr}$. In practice, there is not much physical space in the Recycler Ring for a tuner of this length hence this would be the secondary choice. \label{fig:optimallengths2}}
\end{figure}

There will be two of these `optimal' lengths (one before the avoided crossing - whose position itself is a function of the tuner's length - and one beyond it). Both these lengths were searched for by careful interpolation/extrapolation of results until satisfactory values were found. In both cases, very similar tuning ranges ($\sim$5.5kHz) were found (see Figures \ref{fig:optimallengths} and \ref{fig:optimallengths2}). This draws attention to the idea that the frequency of the accelerating mode has no direct dependence on the length of the tuner. Instead, it effectively only depends on how close it is to the tuner's frequency in ($\mu_\text{ferr}$, l) space.

\label{sec:optimization}

\pagebreak
\section{Double tuner investigation}

\begin{figure}[h]
    \centering
    \includegraphics[width=0.95\textwidth]{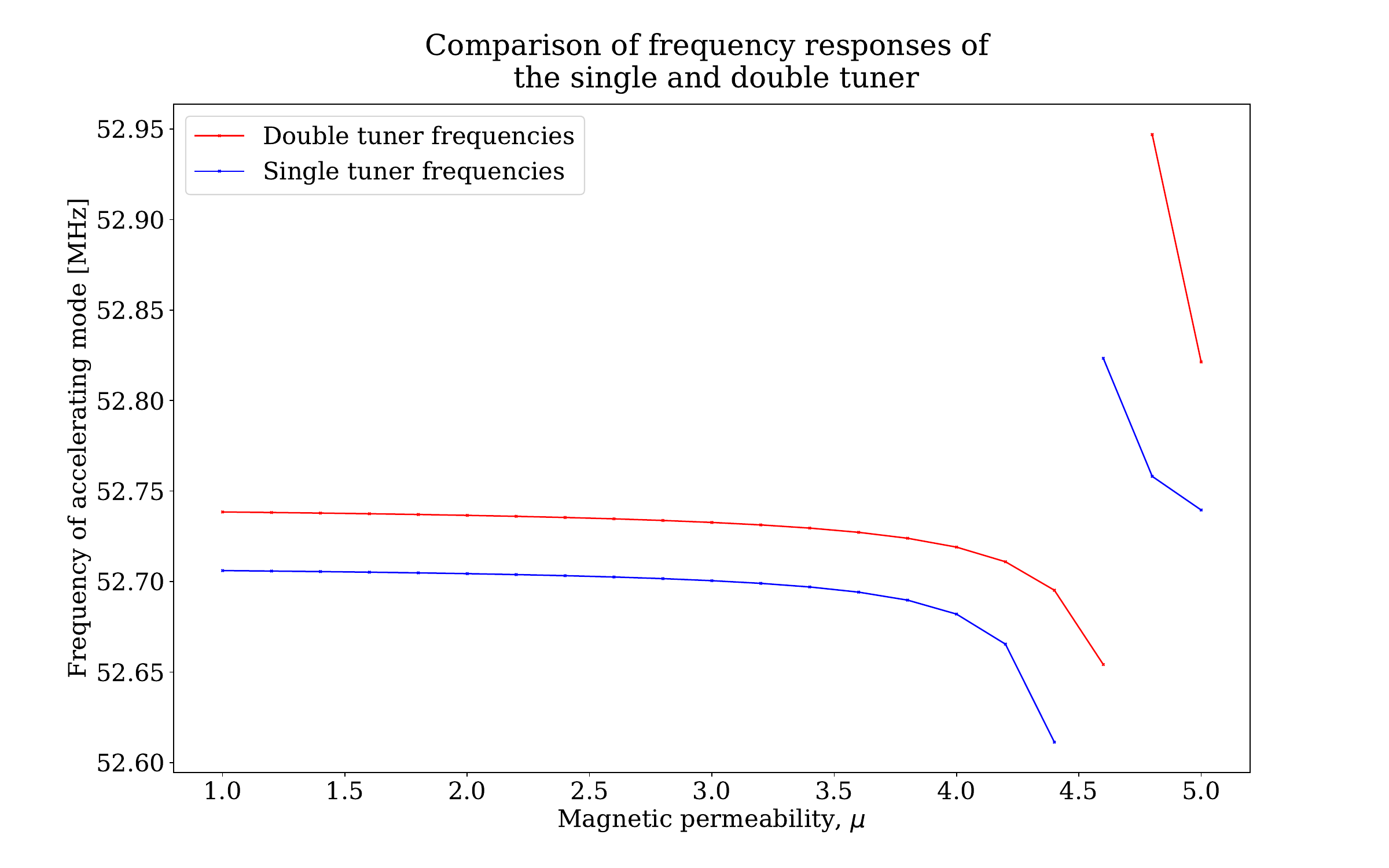}

    \caption{The frequency response of both systems are similar, only with an offset in frequency.\label{fig:singledouble}}
\end{figure}

\begin{figure}[h]
    \centering
    \includegraphics[width=0.95\textwidth]{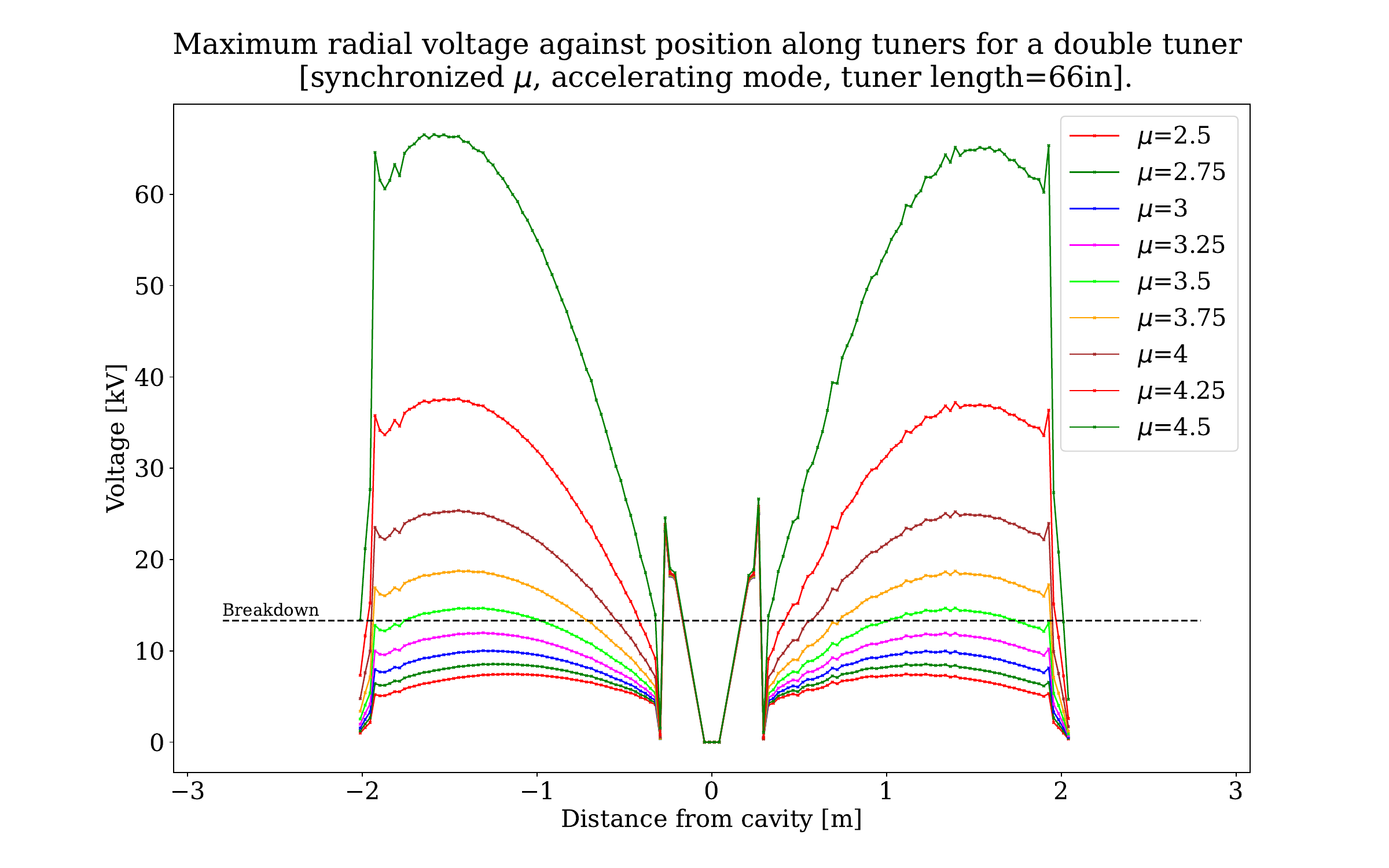}

    \caption{When $\mu_\text{ferr}$'s are synchronized the resonant modes in the tuners are symmetrical as expected. The tuning range is not significantly improved: varying $\mu_\text{ferr}$ between 2 and 3.25 (to remain below breakdown and within the operational $\mu_\text{ferr}$ range) a tuning range just over 4.1kHz is obtained. Expanding the $\mu_\text{ferr}$ range improves the tuning range to $\sim$8kHz which is still below the desired 10kHz. The two peaks near the origin are as expected: they correspond to the electric field amplitude within the cavity (which should go to zero within the inner cylinder of the cavity!).\label{fig:doublesynced}}
\end{figure}

In hopes to increase the tuning range of the cavity in the least costing manner, the frequency of a double tuner cavity was studied. The second tuner was placed exactly opposite the first, on the other side of the main cavity. The frequency-$\mu_\text{ferr}$ graph was very similar to that of a single tuner (see Figure \ref{fig:singledouble}) and very little improvement in the tuning range was seen (see Figure \ref{fig:doublesynced}).

\begin{figure}[h]
    \centering
    \includegraphics[width=0.5\textwidth]{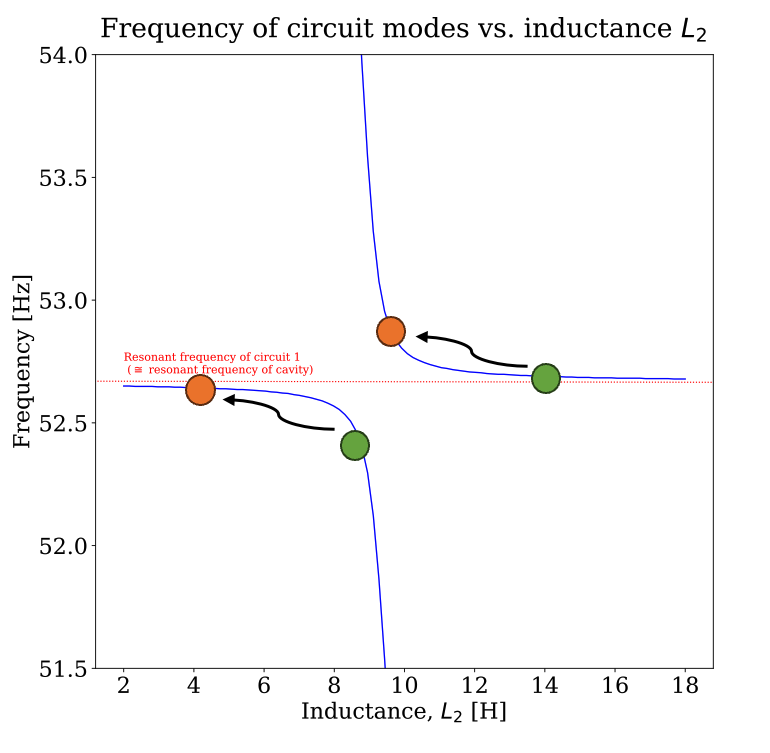}

    \caption{When the system is in its green state the frequency is decreased as much as possible. To get the full frequency shift we change the system to its orange state which theoretically increases the frequency as much as possible. The other tuner should always be far from the crossing otherwise it will change the frequency in the wrong direction (i.e. one tuner increases the frequency and the other decreases it). Ideally, this other tuner should be kept as far away as possible from the crossing. \label{fig:syncexplain}}
\end{figure}

\begin{figure}[h]
    \centering
    \includegraphics[width=0.9\textwidth]{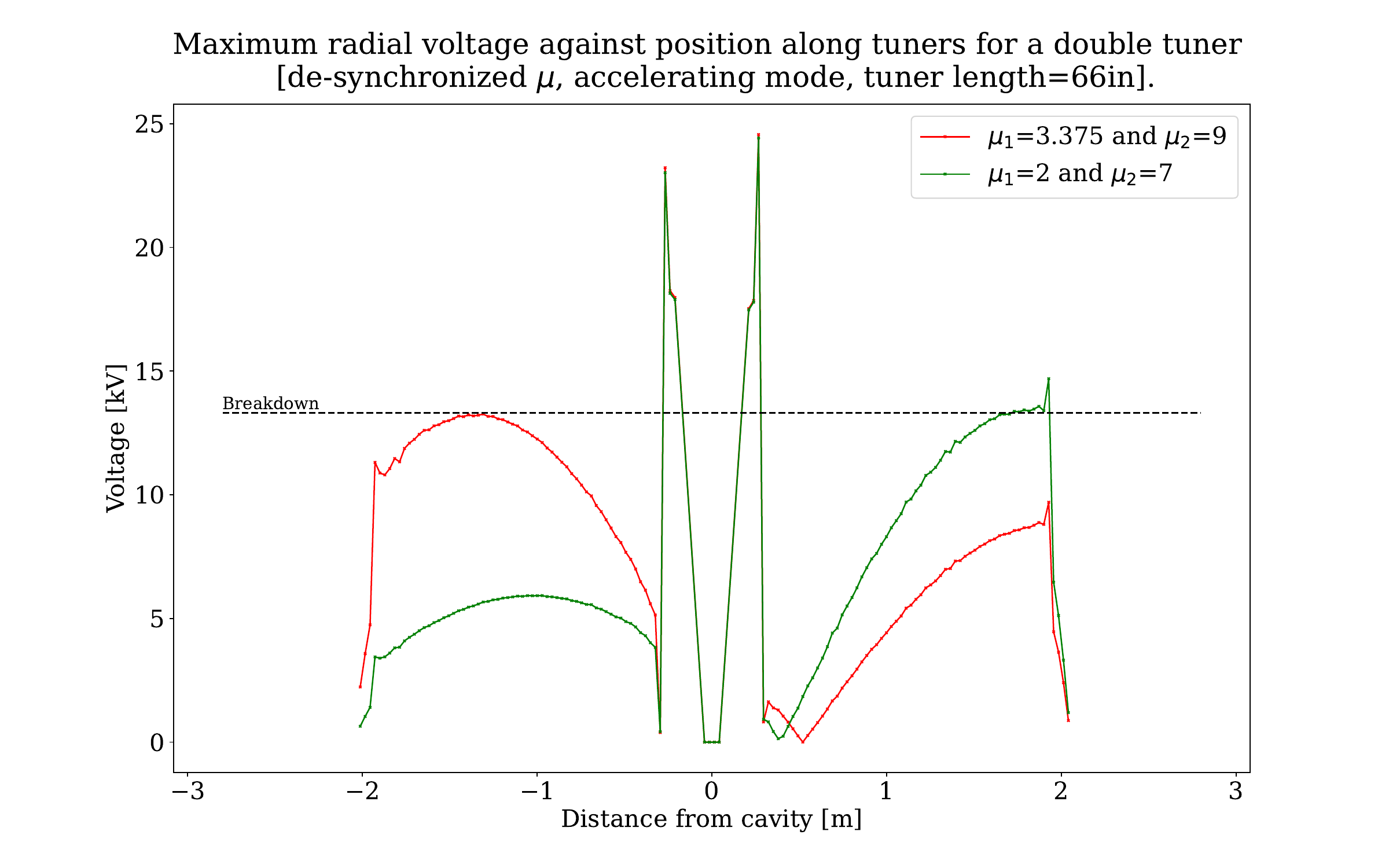}

    \caption{The de-synchronized $\mu$ method allows for more tuning range (=11.025kHz for the $\mu$ values above). \textbf{NB:} In reality this $\mu$ range cannot be used for the Fermilab RR cavities, for which $\mu$ should not be outside of the range $2.5<\mu<4$. Interestingly, the resonant cavity mode seems to be defined across both tuners rather than separately in each.\label{fig:doubleunsynced}}
\end{figure}

In an attempt to exploit the avoided crossing, the $\mu_\text{ferr}$ values of each tuner were de-synchronized. The tuners were both chosen to be of the same length, hence the avoided crossing was at the same value of $\mu_\text{ferr}$ for both. In this set-up, one tuner operates on the \textit{right} of the crossing whilst the other operates on the \textit{left} in $\mu_\text{ferr}$ space. At the first boundary value of the frequency we have (for example) the left tuner as far to the right as possible (hence \textit{decreasing} the frequency as much as possible) whilst the other tuner is kept as far away from the crossing as possible to reduce any counter-acting effects (see the green dots in Figure \ref{fig:syncexplain}). As we shift to the other side of the tuning range, we change the $\mu_\text{ferr}$ accordingly such that the situation is reversed: the latter is increasing the frequency as much as possible whilst the former remains far from the crossing (see the orange dots in the same figure). With this method we can exploit both sides of the avoided crossing and stay below breakdown voltage as required. This proved to be a successful method to increase (though not significantly) the tuning range (see Figure \ref{fig:doubleunsynced}).


\section{Conclusions}

An intuitive explanation behind the tuning range of RF cavities of this kind - strongly supported by predictions from circuit theory - has been outlined. An approximate form of the tuner's resonant frequency was derived. This equation can be used to accurately predict at what length and ferrite magnetic permeability the tuner mode will be met by the resonant frequency: i.e. where there avoided crossing occurs. This equation's form can be re-used for other RF cavities of this type provided data on the tuner's resonant frequency has been gathered. 

Problems linked to the voltage in the tuner were taken into account and a final optimization of the dimensions of the Fermi RR cavities was successfully carried out. Unfortunately, the desired goal of 10kHz of tuning range was not obtained. Nonetheless, having optimized the tuning range as much as possible for these cavities, it has been shown that a tuning range of 10kHz is not attainable for the RF cavity with its current components. The method of optimization for the tuning range, as described in previous sections, can be extended to other/future cavities given access to an EM simulation software.

Simulations of a double tuner were also analysed, although the benefits of this change were often disappointing with rather little increase in tuning range. A new method to maximise the frequency shift by de-synchronizing the $\mu_\text{ferr}$'s of the tuners was proposed. Although more complex, this method gave encouraging results: improving the tuning range by $\sim$3kHz. This idea can be applied to any multiple tuner RF cavity systems. 

The concept of the avoided crossing, ubiquitous in many fields of physics, offers an intuitive explanation behind the frequency shift of the cavity. In experiments involving joint waveguides, the position of the avoided crossing should be carefully studied and simulations should be done to reduce the risk of voltage breakdown in components.

\section{Further investigations}

In future works, the effect of multiple tuner cavities could be further studied. The double tuner investigated in this paper was very symmetric - with the tuners being opposite each other. Systems with tuners at 120$^\circ$ to each other (or higher orders) may yield different, perhaps more interesting, results. Placing the tuners in different places along the cavity may also give useful results. Finally, the `de-synchronized $\mu_\text{ferr}$' method could be improved or implemented in other RF systems.

Moreover, it might be worthwhile to continue studying the resonance mode within the tuner. Its behaviour is not straightforward: for the double tuner case, it even seems to be resonating between both tuners (see Figure \ref{fig:doubleunsynced}). 

\appendix

\acknowledgments
This research used resources of the National Energy Research Scientific Computing Center (NERSC), a U.S. Department of Energy Office of Science User Facility located at Lawrence Berkeley National Laboratory, operated under Contract No. DE-AC02-05CH11231 using NERSC award <HEP>-ERCAP<m349>.\\
Thanks to R. Madrak (FNAL National Laboratory) for excellent assistance on many aspects of this research.




\end{document}